\documentclass[twocolumn,english,prc,superscriptaddress,reprint]{revtex4-1}

\usepackage{amssymb}

\usepackage{float}
\restylefloat{table}
\usepackage{listings}
\usepackage{braket}
\usepackage{amsmath}
\usepackage{textcomp} % for \textonehalf in bibliography
\usepackage{hyperref}

\usepackage{xcolor}

\definecolor{codegreen}{rgb}{0,0.6,0}
\definecolor{codegray}{rgb}{0.5,0.5,0.5}
\definecolor{codepurple}{rgb}{0.58,0,0.82}
\definecolor{backcolour}{rgb}{0.95,0.95,0.92}

\lstdefinestyle{mystyle}{
    backgroundcolor=\color{backcolour},
    commentstyle=\color{codegreen},
    keywordstyle=\color{magenta},
    numberstyle=\tiny\color{codegray},
    stringstyle=\color{codepurple},
    basicstyle=\footnotesize,
    breakatwhitespace=false,
    breaklines=true,
    captionpos=b,
    keepspaces=true,
    numbers=left,
    numbersep=5pt,
    showspaces=false,
    showstringspaces=false,
    showtabs=false,
    tabsize=2,
    xleftmargin=3cm,
    xrightmargin=3cm
}

\usepackage{graphicx}
\usepackage{color}
\begin{document}

\title{NetKet: A Machine Learning Toolkit for Many-Body Quantum Systems}

\author{Giuseppe Carleo}
\affiliation{Center for Computational Quantum Physics, Flatiron Institute, 162 5th Avenue, NY~10010, New York, USA}
%\email{gcarleo@flatironinstitute.org}

% Senior contributors in alphabetic order
\author{Kenny Choo}
\affiliation{Department of Physics, University of Zurich, Winterthurerstrasse 190, 8057~Z\"urich, Switzerland}

\author{Damian Hofmann}
\affiliation{Max Planck Institute for the Structure and Dynamics of Matter, Luruper Chaussee 149, 22761~Hamburg, Germany}

\author{James E. T. Smith}
\affiliation{Department of Chemistry, University of Colorado Boulder, Boulder, Colorado 80302, USA}

\author{Tom Westerhout}
\affiliation{Institute for Molecules and Materials, Radboud University, NL-6525 AJ Nijmegen, The Netherlands}

% Contributors in alphabetic order
\author{Fabien Alet}
\affiliation{Laboratoire de Physique Th\'eorique, IRSAMC, Universit\'e de Toulouse, CNRS, UPS, 31062 Toulouse, France}

\author{Emily J. Davis}
\affiliation{Department of Physics, Stanford University, Stanford, California 94305, USA}

\author{Stavros Efthymiou}
\affiliation{Max-Planck-Institut f\"ur Quantenoptik, Hans-Kopfermann-Stra{\ss}e 1, 85748 Garching bei M\"unchen, Germany}

\author{Ivan Glasser}
\affiliation{Max-Planck-Institut f\"ur Quantenoptik, Hans-Kopfermann-Stra{\ss}e 1, 85748 Garching bei M\"unchen, Germany}

\author{Sheng-Hsuan Lin}
\affiliation{Department of Physics, T42, Technische Universit{\"a}t M{\"u}nchen, James-Franck-Stra{\ss}e 1, 85748 Garching bei M\"unchen, Germany}

\author{Marta Mauri}
\affiliation{Center for Computational Quantum Physics, Flatiron Institute, 162 5th Avenue, NY~10010, New York, USA}
\affiliation{Dipartimento di Fisica, Universit\`a degli Studi di Milano, via Celoria 16, I-20133 Milano, Italy}

\author{Guglielmo Mazzola}
\affiliation{Theoretische Physik, ETH Z\"urich, 8093 Z\"urich, Switzerland}

\author{Christian B. Mendl}
\affiliation{Technische Universit\"at Dresden, Institute of Scientific Computing, Zellescher Weg 12-14, 01069 Dresden, Germany}

\author{Evert van Nieuwenburg}
\affiliation{Institute for Quantum Information and Matter, California Institute of Technology, Pasadena, CA~91125, USA}

\author{Ossian O'Reilly}
\affiliation{Southern California Earthquake Center, University of Southern California, 3651 Trousdale Pkwy, Los Angeles, CA~90089, USA}

\author{Hugo Th\'eveniaut}
\affiliation{Laboratoire de Physique Th\'eorique, IRSAMC, Universit\'e de Toulouse, CNRS, UPS, 31062 Toulouse, France}

\author{Giacomo Torlai}
\affiliation{Center for Computational Quantum Physics, Flatiron Institute, 162 5th Avenue, NY~10010, New York, USA}

\author{Alexander Wietek}
\affiliation{Center for Computational Quantum Physics, Flatiron Institute, 162 5th Avenue, NY~10010, New York, USA}

\begin{abstract}
We introduce NetKet, a comprehensive open source framework for the study of many-body quantum systems using machine learning techniques.
The framework is built around a general and flexible implementation of neural-network quantum states, which are used as a variational ansatz for quantum wavefunctions.
NetKet provides algorithms for several key tasks in quantum many-body physics and quantum technology,
namely quantum state tomography, supervised learning from wavefunction data, and ground state searches for a wide range of customizable lattice models.
Our aim is to provide a common platform for open research and to stimulate the collaborative development of computational methods at the interface of machine learning and many-body physics.
\end{abstract}

\maketitle

\section{Motivation and significance}
\label{sec:motivation}

Recent years have seen a tremendous activity around the development of physics-oriented numerical techniques based on machine learning (ML) tools \cite{carleo_machine_2019}.
In the context of many-body quantum physics, one of the main goals of these approaches is to tackle complex quantum problems using compact representations of many-body states based on artificial neural networks. These representations, dubbed neural-network quantum states (NQS) \cite{carleo_solving_2017}, can be used for several applications.
In the supervised learning setting, they can be used, e.g., to learn existing quantum states for which a non-NQS representation is available \cite{cai_approximating_2018}.
In the unsupervised setting, they can be used to reconstruct complex quantum states from experimental measurements, a task known as quantum state tomography \cite{Torlai2018}.
Finally, in the context of purely variational applications, NQS can be used to find approximate ground- and excited-state solutions of the Schr\"odinger equation \cite{carleo_solving_2017,Choo2018,Glasser2018,Kaubruegger2018,Saito2017,Saito2018}, as well as to describe
unitary \cite{carleo_solving_2017,czischek_quenches_2018,jonsson_neural-network_2018} and dissipative \cite{hartmann_neural-network_2019,yoshioka_constructing_2019,nagy_variational_2019,vicentini_variational_2019}
many-body dynamics.
Despite the increasing methodological and theoretical interest in NQS and their applications, a set of comprehensive, easy-to-use tools for research applications is still lacking. This is particularly pressing as the complexity of NQS-related approaches and algorithms is expected to grow rapidly given these first successes, steepening the learning curve.

The goal of NetKet is to provide a set of primitives and flexible tools to ease the development of cutting-edge ML applications for quantum many-body physics. NetKet also wants to help bridge the gap between the latest and technically demanding developments in the field and those scholars and students who approach the subject for the first time.
Pedagogical tutorials are provided to this aim.
Serving as a common platform for future research, the NetKet project is meant to stimulate the open and easy-to-certify development of new methods and to provide a common set of tools to reproduce published results.

A central philosophy of the NetKet framework is to provide tools that are as simple as possible to use for the end user. Given the huge popularity of the Python programming language and of the many accompanying tools gravitating around the Python ecosystem, we have built NetKet as a full-fledged Python library. This simplicity of use however does not come at the expense of performance. With this efficiency requirement in mind, all critical routines and components of NetKet have been written in C++11.

\section{Software description}
\label{sec:software_description}

We will first give a general overview of the structure of the code in Sect.~\ref{sec:swa} and then provide additional details on the functionality of NetKet in Sect.~\ref{sec: swf}.

\subsection{Software architecture}
\label{sec:swa}

\begin{figure*}[t]
    \centering
    \includegraphics[scale=0.6]{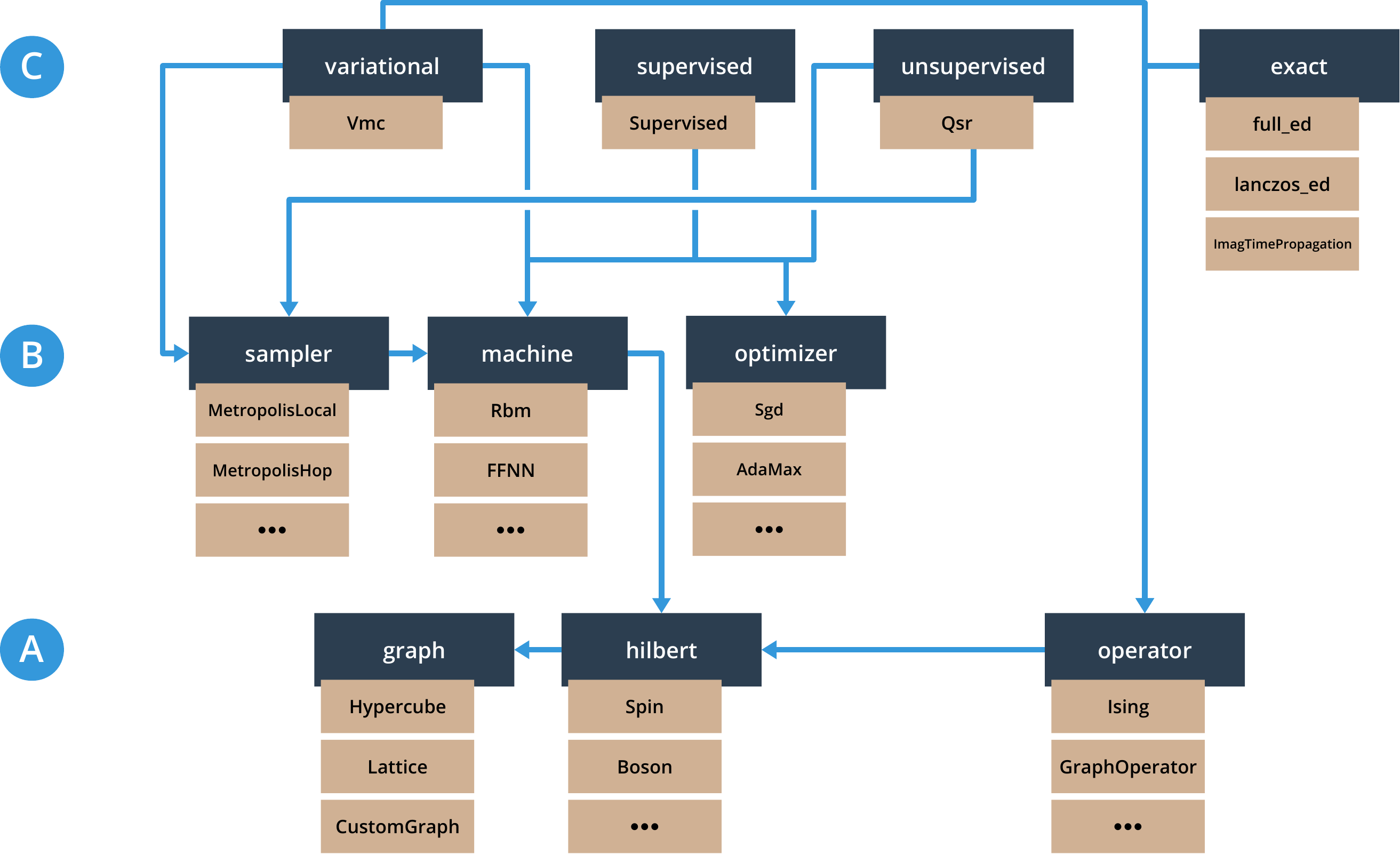}
    \caption{The main submodules of the \texttt{netket} Python module and their dependencies from a user perspective (i.e., only dependencies in the public interface are shown).
    Below each submodule, examples of contained classes and functions are displayed.
    In a typical workflow, users will first define a quantum model (A), specify the desired variational representation of the wavefunction and optimization method (B), and then run the simulation through one of the driver classes (C).
    A more detailed description of the software architecture and features is given in the main text.}
    \label{fig:modules-dependencies}
\end{figure*}

The core of NetKet is implemented in C++.
For ease of use and in order to facilitate the integration with other frameworks, a Python interface is provided, which exposes all high-level functionality from the C++ core via \texttt{pybind11} \cite{pybind11} bindings.
Use of the Python interface is recommended for users building on the library for research purposes, while the C++ code should be modified for extending the NetKet library itself.

NetKet is divided into several submodules.
The modules \texttt{graph}, \texttt{hilbert}, and \texttt{operator} contain the classes necessary for specifying the structure of the many-body Hilbert space, the Hamiltonian, and other observables of a quantum system.

The core component of NetKet is the \texttt{machine} module, which provides different variational representations of the quantum wavefunction, particularly in the form of NQS.
The \texttt{var\-iational}, \texttt{supervised}, and \texttt{unsupervised} modules contain driver classes for energy optimization, supervised learning, and quantum state tomography, respectively.
These driver classes are supported by the \texttt{sampler} and \texttt{optimizer} modules, which provide classes for performing Variational Monte Carlo (VMC) sampling and optimization steps.

The \texttt{exact} module provides functions for exact diagonalization (ED) and imaginary time propagation of the full quantum state,
in order to allow for easy benchmarking and exploration of small systems within the NetKet framework.
ED can be performed by full diagonalization of the Hamiltonian or, alternatively, by a Lanczos procedure, where the user may choose between a sparse matrix representation of the Hamiltonian and a matrix-free implementation. The Lanczos solver is based on the IETL library from the ALPS project \cite{Bauer2011,Albuquerque2007} which implements a variant of the Lanczos algorithm due to Cullum and Willoughby~\cite{Cullum1981,Cullum1985}.
The \texttt{dynamics} module provides a basic Runge-Kutta ODE solver which is used for the exact imaginary time propagation.

The utility modules \texttt{output}, \texttt{stats}, and \texttt{util} contain some additional functionality for output and statistics that is used internally in other parts of NetKet.

An overview of the most important modules and their dependencies is given in Figure~\ref{fig:modules-dependencies}.
A more detailed description of the module contents will be given in the next section.

NetKet uses the Eigen~3 library \cite{eigen3} for linear algebra routines.
In the Python interface, Eigen datatypes are transparently converted to and from NumPy \cite{numpy} arrays by \texttt{pybind11}.
The NetKet driver classes provide methods to directly write the simulation output to JSON files, which is done with the help of the \texttt{nlohmann/json} library for C++ \cite{nlohmann-json}.
Parallelization is implemented based on the Message Passing Interface (MPI), allowing to substantially decrease running time.
Specifically, the Monte Carlo sampling of expectation values implemented in the \texttt{variational.Vmc} class is parallelized, with each node drawing independent samples from the probability distribution which are averaged over all nodes.

\subsection{Software functionalities}
\label{sec: swf}
The core feature of NetKet is the variational representation of quantum states by artificial neural networks.
Given a variational state, the task is to optimize its parameters with regard to a specified loss function, such as the total energy for ground state searches or the (negative) overlap with a given target state.
In this section, we will discuss the models, types of variational wavefunctions, and learning schemes that are available in NetKet.

\subsubsection{Model specification}
\label{sec:model}

NetKet currently supports lattice models with a finite Hilbert space of the form $\mathcal H = \bigotimes_{i=1}^N \mathcal H_\mathrm{local}$
where $N$ denotes the number of lattice sites.
The system is defined on a graph $\mathcal G = (\mathcal V, \mathcal E)$ with a set of sites $\mathcal V$ and a set of edges (also called bonds) $\mathcal E$ between sites.
The graph structure is used to help with the definition of operators on the lattice and to encode the spatial structure of the model, which is necessary, e.g., to work with convolutional neural networks (CNNs).
NetKet provides the predefined \texttt{Hypercube} and \texttt{Lattice} graphs.
Furthermore, \texttt{CustomGraph} supports arbitrary edge-colored graphs, where each edge is associated with an integer label called its color. This color can be used to describe different types of bonds.

Several predefined Hamiltonians are provided, such as spin models (transverse field Ising, Heisenberg models) or bosonic models (Bose-Hubbard model).
%TODO Comment on fermionic models not being available and/or planned for the future?
For specifying other observables and custom Hamiltonians, additional classes are available:
A convenient option for common lattice models is to use the \texttt{GraphOperator} class,
which allows to construct a Hamiltonian from a family of 2-local operators acting on each bond of a selected color and a family of 1-local operators acting on each site.
It is also possible to specify general $k$-local operators (as well as their products and sums) using the \texttt{LocalOperator} class.

\subsubsection{Variational quantum states}
\label{sec:variational}

The purpose of variational states is to provide a compact and computationally efficient representation of quantum states.
Since generally only a subset of the full many-body Hilbert space will be covered by a given variational ansatz, the aim is to use a parametrization that captures the relevant physical states for a given problem.

The variational wavefunctions supported by NetKet are provided as part of the \texttt{machine} module, which currently includes NQS but also Jastrow wavefunctions \cite{RevModPhys.63.1,Becca2017} and matrix-product states (MPS) \cite{White1992,Rommer1997,Schollwck2011}.

Broadly, there are two main types of NQS available in NetKet: restricted Boltzmann machines (RBM) \cite{Hinton2006} and feed-forward neural networks (FFNN) \cite{LeCun2015, Goodfellow2016, Saito2017, Saito2018}.
Both types of networks are fully complex, i.e., with both complex-valued parameters and output.

The \texttt{machine} module contains the \texttt{RbmSpin} class for spin-$\frac{1}{2}$ systems as well as two other variants: the symmetric RBM (\texttt{RbmSpinSymm}) to capture lattice symmetries such as translation and inversion symmetries and the multi-valued RBM (\texttt{RbmMultiVal}) for systems with larger local Hilbert spaces (such as higher spins or bosonic systems).

FFNNs represent a broad and flexible class of networks and are implemented by the \texttt{FFNN} class.
They consist of a sequence of layers available from the \texttt{layer} submodule, each layer performing either an affine transformation to the input vector or applying a non-linear activation function.
There are currently two types of affine maps available:
\begin{itemize}
    \item Dense fully-connected layers, which for an input $\boldsymbol x \in \mathbb{C}^n$ and output $\boldsymbol y \in \mathbb{C}^m$ have the form
    \(
        \boldsymbol y = \mathbf W \boldsymbol x + \boldsymbol b
    \)
    where $\mathbf W \in \mathbb{C}^{m \times n}$ and $\boldsymbol b \in \mathbb C^m$ are called the \emph{weight matrix} and \emph{bias vector}, respectively.
    \item Convolutional layers \cite{Gu2015,LeCun2015} for hypercubic lattices.
\end{itemize}
As activation functions, rectified linear units (\texttt{Relu}) \cite{Nair2010}, hyperbolic tangent (\texttt{Tanh}) \cite{LeCun2012}, and the logarithm of the hyperbolic cosine (\texttt{Lncosh}) are provided.
RBMs without visible bias can be represented as single-layer FFNNs with $\ln\cosh$ activation, allowing for a generalization of these machines to multiple layers \cite{Choo2018}.

Finally, the \texttt{machine} module also provides more traditional variational wavefunctions, namely MPS with periodic boundary conditions (\texttt{MPSPeriodic}) and long-range Jastrow (\texttt{Jastrow}) wavefunctions, which allows for comparison of NQS with results obtained using these approaches.

Custom wavefunctions may be provided by implementing subclasses of the \texttt{AbstractMachine} class in C++.

\subsubsection{Supervised learning}
\label{sec:supervised}

In supervised learning, a target wavefunction is given and the task is to optimize a chosen ansatz to represent it.
This functionality is contained within the \texttt{supervised} module. Given a variational state $\ket{\Psi_{\mathrm{NN}}(\boldsymbol\alpha)}$ depending on the parameters $\boldsymbol{\alpha} \in \mathbb C^m$ and a target state $\ket{\Psi_\mathrm{tar}}$, the negative log overlap
\begin{equation}
\label{eq:overlap-loss}
\mathcal{L}(\boldsymbol{\alpha}) =
    -\log
        \frac{\braket{\Psi_\mathrm{tar}|\Psi_{\mathrm{NN}}(\boldsymbol{\alpha})}}
             {\braket{\Psi_\mathrm{tar}|\Psi_\mathrm{tar}}}
        \frac{\braket{\Psi_{\mathrm{NN}}(\boldsymbol{\alpha})|\Psi_\mathrm{tar}}}
             {\braket{\Psi_{\mathrm{NN}}(\boldsymbol{\alpha})|\Psi_{\mathrm{NN}}(\boldsymbol{\alpha})}}
\end{equation}
is taken as the loss function to be minimized. The loss is computed in a Monte Carlo fashion by direct sampling of the target wavefunction. To minimize the loss, the gradient $\boldsymbol{\nabla}_{\boldsymbol{\alpha}} \,\mathcal{L}$ of the loss function with respect to the parameters is calculated. This gradient is then used to update the parameters according to a specified gradient-based optimization scheme. For example, in stochastic gradient descent (SGD) the parameters are updated as
\begin{equation}
\label{eq:sgd}
\boldsymbol{\alpha} \rightarrow \boldsymbol{\alpha} - \lambda \boldsymbol{\nabla}_{\boldsymbol{\alpha}} \mathcal{L}
\end{equation}
where $\lambda$ is the learning rate.
The different update rules supported by NetKet are contained in the \texttt{optimizer} module. Various types of optimizers are available, including SGD, AdaGrad \cite{Duchi2011}, AdaMax and AdaDelta \cite{Kingma2014}, AMSGrad \cite{Reddi2018}, and RMSProp.

\subsubsection{Unsupervised learning}
NetKet also allows to carry out unsupervised learning of unknown probability distributions, which in this context corresponds to quantum state tomography~\cite{PhysRevA.64.052312}. Given an unknown quantum state, a neural network can be trained on projective measurement data to discover an approximate reconstruction of the state~\cite{Torlai2018}. In NetKet, this functionality is contained within the \texttt{unsupervised.Qsr} class.

For some given target quantum state $|\Psi_\mathrm{tar}\rangle$, the training dataset $\mathcal{D}$ consists of a sequence of projective measurements $\boldsymbol{\sigma}^{\boldsymbol{b}}$ in different bases $\boldsymbol{b}$, with underlying probability distribution $P(\boldsymbol{\sigma}^{\boldsymbol{b}}) = |\Psi_\mathrm{tar}(\boldsymbol{\sigma}^{\boldsymbol{b}})|^2$. The quantum reconstruction of the target state translates into minimizing the statistical divergence between the distribution of the measurement outcomes and the distribution generated by the NQS. This corresponds, up to a constant dataset entropy contribution, to maximizing the log-likelihood of the network distribution over the measurement data
\begin{equation}
\mathcal{L}=\sum_{\boldsymbol{\sigma}^{\boldsymbol{b}}\in\mathcal{D}}\log \pi(\boldsymbol{\sigma}^{\boldsymbol{b}})\:,
\end{equation}
where $\pi$ denotes the probability distribution
\begin{equation}
    \pi(\boldsymbol{\sigma}) = \frac{ \left| \Psi_\textrm{NN}(\boldsymbol{\sigma})\right|^{2}}{\sum_{\boldsymbol{\sigma}'} \left| \Psi_\textrm{NN}(\boldsymbol{\sigma}')\right|^{2}}.
    \label{Eq::NQS_distribution}
\end{equation}
generated by the NQS wavefunction.

Note that, for every training sample where the measurement basis differs from the reference basis $|\boldsymbol{\sigma}\rangle$ of the NQS, a unitary transformation $\hat{U}$ should be applied to appropriately change the basis, $\Psi_{\mathrm{NN}}(\boldsymbol{\sigma}^{\boldsymbol{b}})=\hat{U}_{\boldsymbol{b}}\Psi_{\mathrm{NN}}(\boldsymbol{\sigma})$.

The network parameters are updated according to the gradient of the log-likelihood $\mathcal{L}$. This can be computed analytically, and it requires expectation values over both the  training data points and the network distribution $ \pi(\boldsymbol{\sigma})$. While the first is trivial to compute, the latter should be approximated by a Monte Carlo average over configurations sampled from a Markov chain.

\begin{figure}[t]
\centering
    \includegraphics[width=1.0\columnwidth]{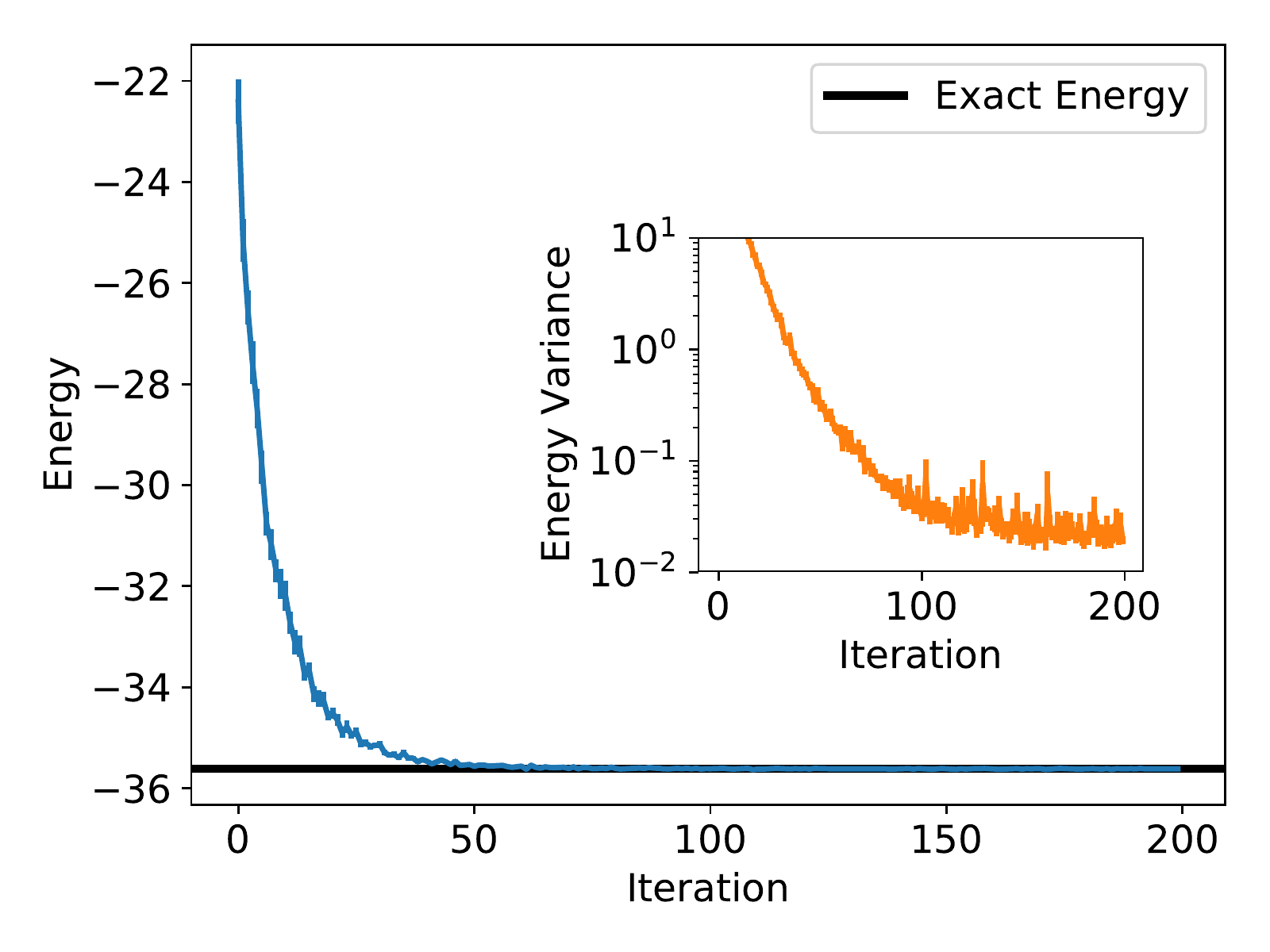}
    \caption{Variational optimization of the restricted Boltzmann machine for the one-dimensional spin-$\frac{1}{2}$ Heisenberg model. The main plot shows the Monte Carlo energy estimate, which converges to the exact ground state energy up to a relative error $\left|(E - E_{\mathrm{exact}})/E_{\mathrm{exact}}\right|$ of $4.16\times 10^{-5}$ within the 200 iteration steps shown. The inset shows the Monte Carlo estimate of the energy variance, which becomes zero in an exact eigenstate of the Hamiltonian.}
\label{fig: Heisenberg}
\end{figure}

\begin{figure}[t]
\centering
    \includegraphics[width=1.0\columnwidth]{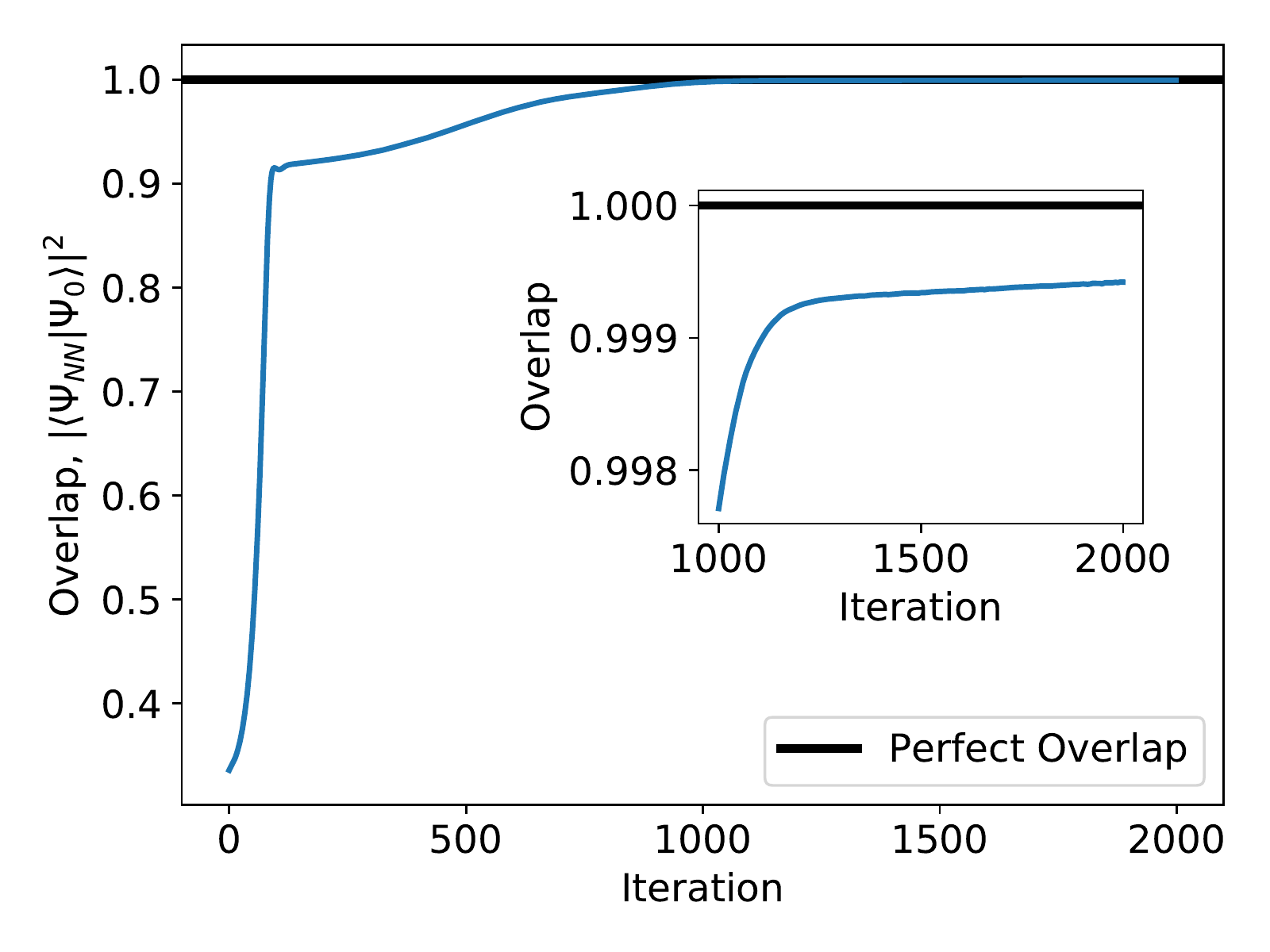}
    \caption{Supervised learning of the ground state of the one-dimensional spin-$\frac{1}{2}$ transverse field Ising model with 10 sites from ED data, using an RBM with $20$ hidden units.
    The blue line shows the overlap between the RBM wavefunction and the exact wavefunction for each iteration.}
\label{fig: Supervised}
\end{figure}

\subsubsection{Variational Monte Carlo}
Finally, NetKet supports ground state searches for a given many-body quantum Hamiltonian $\hat{H}$. In this context, the task is to optimize the parameters of a variational wavefunction $\Psi$ in order to minimize the energy $\langle\hat{H}\rangle$.
The \texttt{variational.Vmc} driver class contains the main logic to optimize a variational wavefunction given a Hamiltonian, a sampler, and an optimizer.

The energy of a wavefunction $\Psi(\boldsymbol{\sigma}) = \braket{\boldsymbol{\sigma} | \Psi}$ can be estimated as
\begin{equation}
\begin{split}
\langle \hat{H} \rangle &= \frac{\sum_{\boldsymbol{\sigma}, \boldsymbol{\sigma}'} \Psi^*(\boldsymbol{\sigma})\bra{\boldsymbol{\sigma}}\hat{H}\ket{\boldsymbol{\sigma}'}\Psi (\boldsymbol{\sigma}')}{\sum_{\boldsymbol{\sigma}} \left| \Psi(\boldsymbol{\sigma})\right|^{2}} \\
&= \sum_{\boldsymbol{\sigma}} \left( \sum_{\boldsymbol{\sigma}'} \bra{\boldsymbol{\sigma}}\hat{H}\ket{\boldsymbol{\sigma}'} \frac{\Psi(\boldsymbol{\sigma}')}{\Psi(\boldsymbol{\sigma})} \right) \frac{ \left| \Psi(\boldsymbol{\sigma})\right|^{2}}{\sum_{\boldsymbol{\sigma}'} \left| \Psi(\boldsymbol{\sigma}')\right|^{2}} \\
&\approx \left\langle \sum_{\boldsymbol{\sigma}'} \bra{\boldsymbol{\sigma}}\hat{H}\ket{\boldsymbol{\sigma}'} \frac{\Psi(\boldsymbol{\sigma}')}{\Psi(\boldsymbol{\sigma})} \right\rangle_{\boldsymbol\sigma}
\end{split}
\end{equation}
where in the last line $\left\langle {}\cdot{} \right\rangle_{\boldsymbol\sigma}$ denotes a stochastic expectation value taken over a sample of configurations $\{\boldsymbol{\sigma}\}$ drawn from the probability distribution corresponding to the variational wavefunction \eqref{Eq::NQS_distribution}.
This sampling is performed by classes from the \texttt{sampler} module, which generate Markov chains of configurations using the Metropolis algorithm \cite{Metropolis1953} to ensure detailed balance.
Parallel tempering \cite{Swendsen1986} options are also available to improve sampling efficiency.

In order to optimize the parameters of a machine to minimize the energy, a gradient-based optimization scheme can be applied as discussed in the previous section.
The energy gradient can be estimated at the same time as $\braket{\hat H}$ \cite{Becca2017,carleo_solving_2017}.
This requires computing the partial derivatives of the wavefunction with respect to the variational parameters, which can be obtained analytically for the RBM \cite{carleo_solving_2017} or via backpropagation \cite{LeCun2015,LeCun2012,Goodfellow2016} for multi-layer FFNNs.
In this case, the steepest descent update according to Eq.~\eqref{eq:sgd} is also a form of SGD, because the energy is estimated using a subset of the full data available from the variational wavefunction.
Alternatively, often more stable convergence can be achieved by using the stochastic reconfiguration (SR) method \cite{Sorella_SR,casula}, which approximates the imaginary time evolution of the system on the submanifold of variational states.
The SR approach is closely related to the natural gradient descent method used in machine learning \cite{Amari1998}. In the NetKet implementation, SR is performed using either an exact or an iterative linear solver, the latter being recommended when the number of variational parameters is large.

Information on the optimization run (sampler acceptance rates, energy, energy variance, expectation of additional observables, and the current variational parameters) for each iteration can be written to a log file in JSON format.
Alternatively, they can be accessed directly inside the simulation loop in Python to allow for more flexible output.

\begin{lstlisting}[
language=Python,
float=*t,
caption=Example script for finding the ground state of the one-dimensional spin-$\frac{1}{2}$ Heisenberg model using an RBM ansatz.,
captionpos=b,
label={lst: VMC}]
import netket as nk

# Define the graph: a 1D chain of 20 sites with periodic
# boundary conditions
g = nk.graph.Hypercube(length=20, n_dim=1, pbc=True)

# Define the Hilbert Space: spin-half degree of freedom at each
# site of the graph, restricted to the zero magnetization sector
hi = nk.hilbert.Spin(s=0.5, total_sz=0.0, graph=g)

# Define the Hamiltonian: spin-half Heisenberg model
ha = nk.operator.Heisenberg(hilbert=hi)

# Define the ansatz: Restricted Boltzmann machine
# with 20 hidden units
ma = nk.machine.RbmSpin(hilbert=hi, n_hidden=20)

# Initialise with machine parameters
ma.init_random_parameters(seed=1234, sigma=0.01)

# Define the Sampler: metropolis sampler with local
# exchange moves, i.e. nearest neighbour spin swaps
# which preserve the total magnetization
sa = nk.sampler.MetropolisExchange(graph=g, machine=ma)

# Define the optimiser: Stochastic gradient descent with
# learning rate 0.01.
opt = nk.optimizer.Sgd(learning_rate=0.01)

# Define the VMC object: Stochastic Reconfiguration "Sr" is used
gs = nk.variational.Vmc(hamiltonian=ha, sampler=sa,
                        optimizer=opt, n_samples=1000,
                        use_iterative=True, method='Sr')

# Run the VMC simulation for 1000 iterations
# and save the output into files with prefix "test"
# The machine parameters are stored in "test.wf"
# while the measurements are stored in "test.log"
gs.run(output_prefix='test', n_iter=1000)
\end{lstlisting}

\begin{lstlisting}[
language=Python,
float=*t,
caption=Example script for supervised learning. A RBM ansatz is optimized to represent the ground state of the one-dimensional spin-$\frac{1}{2}$ transverse field Ising model obtained by ED for this example.,
captionpos=b,
label={lst: Supervised}]
import netket as nk
from numpy import log

# 1D Lattice
g = nk.graph.Hypercube(length=10, n_dim=1, pbc=True)

# Hilbert space of spins on the graph
hi = nk.hilbert.Spin(s=0.5, graph=g)

# Ising spin Hamiltonian
ha = nk.operator.Ising(h=1.0, hilbert=hi)

# Perform Exact Diagonalization to get lowest eigenvector
res = nk.exact.lanczos_ed(ha, first_n=1, compute_eigenvectors=True)

# Store eigenvector as a list of training samples and targets
# The samples would be the Hilbert space configurations and
# the targets should be wavefunction amplitudes.
hind = nk.hilbert.HilbertIndex(hi)
h_size = hind.n_states
targets = [[log(res.eigenvectors[0][i])] for i in range(h_size)]
samples = [hind.number_to_state(i) for i in range(h_size)]

# Machine: Restricted Boltzmann machine
# with 20 hidden units
ma = nk.machine.RbmSpin(hilbert=hi, n_hidden=20)
ma.init_random_parameters(seed=1234, sigma=0.01)

# Optimizer
op = nk.optimizer.AdaMax()

# Supervised Learning module
spvsd = nk.supervised.Supervised(machine=ma,
                                 optimizer=op,
                                 batch_size=400,
                                 samples=samples,
                                 targets=targets)

# Run the optimization for 2000 iterations
spvsd.run(n_iter=2000, output_prefix='test',
	 	  		loss_function="Overlap_phi")
\end{lstlisting}

\section{Illustrative examples}
\label{sec:examples}

NetKet is available as a Python package and can be obtained from the Python package index (PyPI) \cite{PyPI}.
Assuming a properly configured Python environment, NetKet can be installed via the shell command
\begin{verbatim}
    pip install netket
\end{verbatim}
which will download, compile, and install the package.
A working MPI environment is required to run NetKet.
In case multiple MPI installations are present on the system and in order to avoid potential conflicts, we recommend to run the installation command as
\begin{verbatim}
    CC=mpicc CXX=mpicxx pip install netket
\end{verbatim}
with the desired MPI environment loaded in order to perform the build with the correct compiler.
After a successful installation, the NetKet module can be imported in Python scripts.

Alternatively to installing NetKet locally, NetKet also uses the deployment of BinderHub from mybinder.org \cite{Jupyter2018} to build and deploy a stable version of the software, which can be found at \url{https://mybinder.org/v2/gh/netket/netket/master}. This allows users to run the tutorials or other small jobs without installing NetKet.

\subsection{One-dimensional Heisenberg model}
As a first example, we present a Python script for obtaining a variational RBM representation of the ground state of the spin-$\frac{1}{2}$ Heisenberg model on a one-dimensional chain with periodic boundary conditions.
The code for this example is shown in Listing~\ref{lst: VMC}.
Figure \ref{fig: Heisenberg} shows the evolution of the energy expectation value over the course of the optimization run.
We see that for a small chain of 20 sites and an RBM with 20 hidden units, the energy converges to a relative error of the order $10^{-5}$ within about 100 iteration steps.

\subsection{Supervised learning}
As a second example, we use the supervised learning module in NetKet to optimize an RBM to represent the ground state of the transverse field Ising model.
The example script is shown in Listing~\ref{lst: Supervised}.
The exact ground state wavefunction is first obtained by exact diagonalization and then used for training the RBM state by minimizing the overlap loss \eqref{eq:overlap-loss}.
%Mention applications where supervised learning could be useful (when you don't have the exact wave-function)?
Figure~\ref{fig: Supervised} shows the evolution of the overlap over the training iterations.

\section{Impact}
\label{sec:impact}

Given the flexibility of NetKet, we envision several potential applications of this library both in data-driven experimental research and in more theoretical, problem-driven research on interacting quantum many-body systems.
For example, several important theoretical and practical questions concerning the expressibility of NQS, the learnability of experimental quantum states,
and the efficiency at finding ground states of $k$-local Hamiltonians, can be directly addressed using the current functionality of the software.

Moreover, having an easy-to-extend set of tools to work with NQS-based applications can propel future research in the field, without researchers having
to pay a significant cost of entry in terms of algorithm implementation and testing. Since its early release in April 2017, NetKet has already been used for research purposes by several groups worldwide.
% citations to these papers?
We also hope that, building upon a common set of tools, practices like publishing accompanying codes to research papers,
largely popular in the ML community, can become standard practice also for ML applications in quantum physics.

Finally, for a fast-growing community like ML for quantum science, it is also crucial to have
pedagogical tools available that can be conveniently used by new generations of students and researchers.
Benefiting from a growing set of tutorials and step-by-step explanations, NetKet can be comfortably used in schools and lectures.

\section{Conclusions and future directions}
\label{sec:conclusions}

We have introduced NetKet, a comprehensive open source framework for the study of many-body quantum systems using machine learning techniques.
Central to this framework are variational parameterizations of many-body wavefunctions in the form of artificial neural networks.
NetKet is a Python framework implemented in C++11, designed with efficiency as well as ease of use in mind. Several examples, tutorials, and notebooks are provided with our software in order to reduce the learning curve for newcomers.

The NetKet project is meant to continuously evolve in future releases, welcoming suggestions and contributions from its users. For example, future versions may provide a natural interface with general ML frameworks such as PyTorch \cite{paszke2017automatic} and Tensorflow \cite{tensorflow2015-whitepaper}. On the algorithmic side, future goals include the extension of NetKet to incorporate unitary dynamics \cite{carleo_localization_2012,jonsson_neural-network_2018} as well as support for neural density matrices \cite{torlai_latent_2018}.

\section*{Acknowledgements}
\label{sec:thanks}

We acknowledge support from the Flatiron Institute of the Simons Foundation.
J.E.T.S. gratefully acknowledges support from a fellowship through The Molecular Sciences Software Institute under NSF Grant ACI1547580.
H.T. is supported by a grant from the Fondation CFM pour la Recherche.
S.E. and I.G. are supported by an ERC Advanced Grant QENOCOBA under the EU Horizon2020 program (grant agreement 742102) and the German Research Foundation (DFG) under Germany's Excellence Strategy through Project No. EXC-2111 - 390814868 (MCQST).
This project makes use of other open source software, namely pybind11 \cite{pybind11}, Eigen \cite{eigen3}, ALPS IETL \cite{Bauer2011, Albuquerque2007}, nlohmann/json \cite{nlohmann-json}, and NumPy \cite{numpy}.
We further acknowledge discussions with, as well as bug reports, comments, and more from S.~Arnold, A.~Booth, A.~Borin, J.~Carrasquilla, S.~Lederer, Y.~Levine, T.~Neupert, O.~Parcollet, A.~Rubio, M. A.~Sentef, O.~Sharir, M.~Stoudenmire, and N.~Wies.

\section*{References}
\bibliographystyle{elsarticle-num}
%\biboptions{sort&compress}
\interlinepenalty=10000
\bibliography{biblio.bib}



%% References:


%Please add the reference to the software repository if DOI for software is available.

\end{document}